\documentclass[aps,prl,reprint,superscriptaddress,notitlepage]{revtex4-2}

\usepackage{graphicx}% Include figure files
\usepackage{dcolumn}% Align table columns on decimal point
\usepackage{bm}% bold math
\usepackage[version=4]{mhchem}
\usepackage{blindtext}
\usepackage{upgreek}
\usepackage{xcolor}
\usepackage{soul}
\usepackage{ulem}

\usepackage{lineno}
\newcommand{\cdas}{Cd${}_3$As${}_2$}
\newcommand{\zrte}{ZrTe${}_5$}
\newcommand{\hfte}{HfTe${}_5$}

\newcommand{\tth}{$\tan\theta_H$}

\newcommand{\rxx}{$\rho_{xx}$}
\newcommand{\rxy}{$\rho_{xy}$}

\newcommand{\cm}{cm${}^{-3}$}

\begin{document}

\title{Interplay of Quasi-Quantum Hall Effect and Coulomb Disorder in Semimetals}
	\author{Corresponding Author: Ian A. Leahy}
    \email{Corresponding Author: Ian.Leahy@nrel.gov}
	\affiliation{National Laboratory of the Rockies, Golden, Colorado 80401, USA}	
	  \author{Anthony D. Rice}
	\affiliation{National Laboratory of the Rockies, Golden, Colorado 80401, USA}
	\author{Jocienne N. Nelson}
         \altaffiliation{{\it Current Affiliation: First Solar Inc., 1035 Walsh Ave., Santa Clara, 			95050, 			CA, United States}}
	\affiliation{National Laboratory of the Rockies, Golden, Colorado 80401, USA}
    \author{Herve Ness}
    \affiliation{Department of Physics, King's College London, Strand, London WC2R 2LS, United Kingdom}
    \author{Mark van Schilfgaarde}
    \affiliation{National Laboratory of the Rockies, Golden, Colorado 80401, USA}
    \author{David Graf}
    \affiliation{National High Magnetic Field Laboratory, Tallahassee, Florida 32310, USA}
    \author{Alexey Suslov}
    \affiliation{National High Magnetic Field Laboratory, Tallahassee, Florida 32310, USA}
    \author{Wei Pan}
    \affiliation{Materials Physics Department, Sandia National Laboratories, Livermore, California 94551, USA}
    \author{Kirstin Alberi}
	\affiliation{National Laboratory of the Rockies, Golden, Colorado 80401, USA}
	\affiliation{Renewable and Sustainable Energy Institute, National Renewable Energy 					Laboratory and University of Colorado, Boulder, 80301, CO, United States}

	\begin{abstract}
       Low carrier densities in topological semimetals (TSMs) enable the exploration of novel magnetotransport in the quantum limit (QL). Recent findings consistent with 3D quasi-quantum Hall effect (QQHE) have positioned TSMs as promising platforms for exploring 3D quantum Hall transport, but the lack of tunability in the Fermi level has thus far limited the ability to observe a QQHE signal. Here, we tune the defect concentrations in the Dirac semimetal \cdas{} to achieve ultra-low carrier concentrations at 2 K around $2.9\times10^{16}$cm${}^{-3}$, giving way to QQHE signal at modest fields near 10 T. At low carrier densities, where QQHE is most accessible, we find that clear QQHE is obscured by a carrier density dependent background originating from Coulomb disorder from charged point defects and Landau level broadening. Our results highlight the interplay between QQHE and Coulomb disorder, demonstrating that clear observation of QQHE in TSMs intricately depends on Fermi level and disorder magnitudes. We find that Coulomb disorder, as theoretically predicted, is an essential ingredient for understanding the magnetoresistivity for a spectrum of Fermi levels in \cdas{},  anchoring the role of defects and charged disorder in TSM applications. We discuss future constraints and opportunities in exploring 3D QQHE and quantum Hall effects in TSMs.
        \end{abstract}
  	
  	\maketitle


\section{Introduction}
The 2D quantum Hall effect (QHE) is a hallmark of two-dimensional electron gases (2DEG) in magnetic fields, resulting in quantized plateaus in the Hall resistivity \rxy{} paired with vanishingly small longitudinal resistivity, \rxx{} \cite{Klitzing1980,Klitzing1986}. This quantized effect, observable across a spectrum of material systems regardless of chemical composition or the presence of reasonable disorder, is accompanied by dissipationless chiral edge states \cite{Laughlin1981,Halperin1982}. The 2D QHE has sparked decades of research on related effects, enabled high fidelity measurements of fundamental constants ($h/e^2$), generated many potential applications in spintronics and quantum computing \cite{Yu2018,LiSM2018}, and critically, has helped elucidate the role of topology in the study of electronic band structures. Observation of a 3D QHE stands to have a similar impact, though experimental confirmation has remained elusive. In recent years the 3D quasi-quantum Hall effect (QQHE), an analog of the 2D QHE, has been theoretically predicted and experimentally observed in a few low-carrier density topological semimetal (TSM) systems, opening the door to explore new quantum Hall phases \cite{Tang2019,Galeski2021,Gooth2023,Manna2022,Wawrzynczak2022,Galeski2020,Piva2024}.Though 3D QQHE systems lack the topological protection of true 3D QHE states, they may offer a route towards new correlated bulk states and 3D QHE-like surface states \cite{Gooth2023}. As in 2D, disorder is necessary for observing QQHE, but we expect it to play a significant obscuring role in the 3D QQHE as compared to the 2D QHE due to the lack of a bulk gap and dispersing Landau bands. This work marks a first attempt to understand how disorder impacts the observation of QQHE magnetotransport in TSMs.

Herein, we study the effects of Coulomb disorder on QQHE and electrical transport in ultra-low carrier density \cdas{} bulk-like thin films. The magnetotransport is described as a superposition of the QQHE and Coulomb disorder scattering. We demonstrate a quantitative analysis for understanding the relative QQHE and Coulomb disorder scattering contributions to \rxx{}. We find that Coulomb disorder in low carrier density (Dirac) semimetals obscures the QQHE in two main ways: i) smoothing features by broadening Landau bands (LBs) and ii) affecting quasi-quantization of \rxy{} by lowering mobility and introducing a field-dependent Coulomb scattering contribution to the magnetoconductivity, spoiling the approximation $\rho_{xy}\approx\sigma_{xy}^{-1}$. We highlight an intertwined balance between low carrier densities and Coulomb disorder effects that should be optimized in pursuit of QQHE in low density semimetals. QQHE signatures are most visible near the QL. Low carrier densities enable lower QL magnetic fields. Reducing the carrier density in semimetals increases the Coulomb disorder potential, resulting in broadened LBs and increased, field-dependent, scattering. Our observations generate actionable insight into the exploration and enhancement of QQHE in low-density semimetals broadly. 

\begin{figure*}[ht!]
    \centering
    \includegraphics[trim={.00cm 0cm 0cm 7.1cm},width = .85\linewidth,clip]{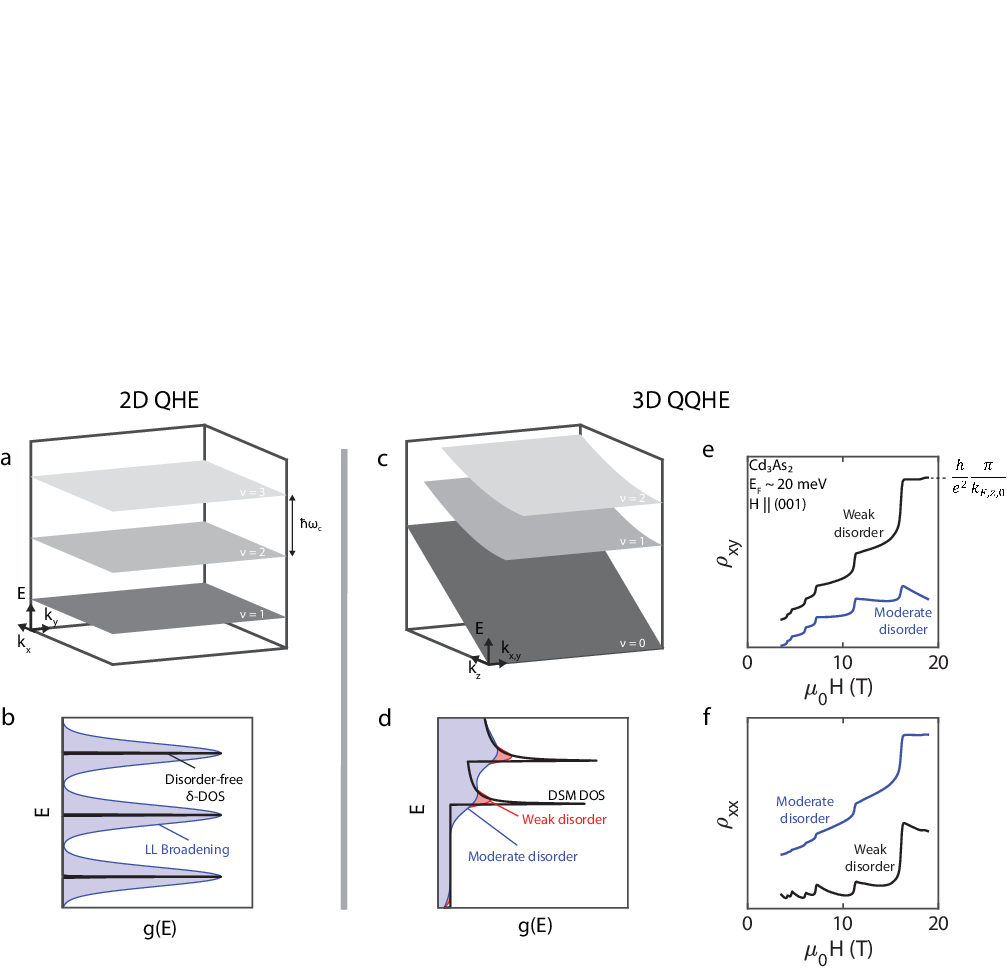}
    \caption{{\bf Comparison of 2D QHE and 3D QQHE.} Schematic band structures and density of states (DOS) for a field applied along the z-direction for a,b) a 2D QHE system and c,d) a 3D QQHE system with Dirac dispersion. In the 2D QHE, LLs resulting from the applied field give a delta-function DOS which broadens with disorder. Changing the applied field moves LLs through the $E_F$, resulting in Hall plateaus. c) Landau bands for a 3D DSM, showing the remaining dispersion of the bands along the applied field direction. The DOS does not drop to zero between Landau bands because of this dispersion. As in 2D, disorder broadens the 3D DSM DOS. e) Hall and f) longitudinal resistivity calculated from a $k \cdot p$ model for 3D DSM \cdas{} in the case of weak (black traces) and moderate (blue traces, offset) disorder for field along the c-axis. The chemical potential is fixed at $20$ meV. $\rho_{xx}(H)$ exhibits peak-like features, similar to the 2D QHE (\rxx{} traces scaled for clarity). When only the lowest LB is occupied, \rxy{} is quasi-quantized at a value contingent on the Fermi wave vector along the applied field direction, $k_{F,c}$. Moderate disorder introduces mixing of the longitudinal and Hall conductivities which affects the quasi-quantization of \rxy{}. These calculations do not account for the effects of LB broadening on the DOS. For a full discussion of theoretical calculations, disorder treatment, and comparison to the featureless case of fixed carrier density, see Methods and Supplementary Note 1. }
    \label{fig:fig1s1}
\end{figure*}

\vspace{1cm}
To understand the 3D QQHE, we begin with a description of the 2D QHE. The quantized Hall conductivity in the 2D QHE arises from the interplay of dimensionality, Landau levels (LLs), and disorder. A weakly disordered 2DEG in a magnetic field undergoes Landau quantization as shown in Fig. 1a, which quantizes the band structure into degenerate and non-dispersive levels. This leads to a density of states (DOS) comprised of delta functions separated in energy by $\hbar\omega_c$ (Fig. 1b), where $\omega_c=eB/m_\mathrm{eff}$ is the cyclotron frequency. As the magnetic field is increased, LLs move through the Fermi level, resulting in a finite then vanishing DOS. Charged point defects broaden the LL $\delta$-DOS into a distribution of states. Extended states carry current across the 2DEG while localized states form at charged defect sites \cite{Cooper1993,Stormer1998}. Disorder is necessary for observed 2D QHE: localization at charged defects gives the plateaux finite width. The value of the quantized Hall plateaus corresponds to the number of filled LLs.  Due to the topological nature of the QHE, the quantization of $\sigma_{xy}$ is tolerant of disorder as long as the disorder is not strong enough to cause LLs to blur into one another. 

\begin{figure*}
    \centering
    \includegraphics[width=0.85\linewidth]{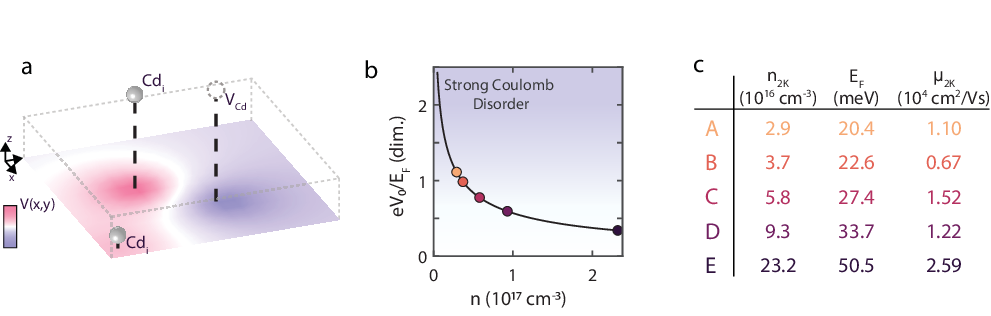}
    \caption{{\bf Coulomb disorder in \cdas{}}. a) Charged Cd interstitials and vacancies have concentrations of $\sim10^{18}$\cm{} in \cdas{} films. Paired with Fermi-level dependent screening, they generate a disorder potential. The surface plot shows a slice of the disorder potential, $V(x,y)$, at the plane defined by the dashed line. b) Ratio of the average disorder potential strength to the Fermi level as a function of $n$, calculated using Ref. \cite{Skinner2014} and relevant values for \cdas{} (see Supplementary Note 5). Colored circles represent estimates for Samples A-E presented here. As the carrier density decreases, \cdas{} moves into the strong Coulomb disorder limit. c) Summary of carrier densities, Fermi levels, and mobilities for Samples A-E. Carrier densities extracted from the low field slope of $\rho_{xy}(H)$ are used in combination with the DOS to estimate $E_F$ (See Supplementary Note 1).}
    \label{fig:fig1s2}
\end{figure*}

The concept of the QHE has long been extended to 3D systems, though experimental observations are yet to confirm a 3D QHE \cite{Halperin1987,Cooper1989,Kohmoto1992,Koshino2001,Koshino2003}. Like 2D QHE, true 3D QHE characterized by vanishing longitudinal resistivity and precisely quantized Hall plateaus, requires a bulk gap \cite{Halperin1987,Kohmoto1992,Gooth2023}. Recent experimental signatures of a quasi-quantized Hall effect in gapless TSM systems such as \zrte{} or \hfte{} have reinvigorated the field of 3D quantum Hall physics. These studies reveal observable, but not precisely quantized, plateaus in $\rho_{xy}$ accompanied by correlated, but non-zero, minima in $\rho_{xx}$. When only the lowest Landau band (LB) is occupied in the quantum limit (QL), the Hall resistivity reaches a final plateau value of $\rho_{xy}^{FP}=\pi h/e^2k_{F,z,0}$, where $k_{F,z,0}$ is the Fermi momentum along the applied field direction of the lowest LB. These transport behaviors, dubbed the QQHE, are observable in 3D systems and have been predicted to be the behavioral rule in low-carrier density metals \cite{Tang2019,Galeski2021,Gooth2023,Manna2022,Wawrzynczak2022,Galeski2020,Piva2024}. The 3D QQHE is manifestly not the same as a true 3D QHE: plateaus are not explicitly quantized, and the states are not topologically robust, though unique surface states remain.

To highlight the mechanistic origins of the QQHE and the differences to a true 3D QHE, we consider a simplified band structure for a linearly-dispersing 3D Dirac semimetal (DSM) with an applied magnetic field along the z-direction in Fig. 1c. The resulting DOS is depicted in Fig. 1d. Landau quantization now generates LBs which are dispersive. The 3D DSM DOS shows sharp delta-function-like features superimposed on a finite background which grows in energy reflective of the dispersing LBs. As in quasi-2D quantum Hall systems, the Hall conductivity can be expressed as an integral over $k_z$ of $k_x$-$k_y$ Hall conductivity slices. Recent work has found that the total Hall conductivity is given as $\sigma_{xy}=2e^2/h\sum_{\mathrm{Occ, }\nu} k_{F,z,\nu}$, similar to a 2D system with an added dependence on the Fermi wavevector \cite{Gooth2023}. Experimentally, the plateau-like features observed in systems like \zrte{} evolve from standard Shubnikov-de-Haas oscillations and are thus a generic result of having few LBs populated that are well separated in energy.

As the applied magnetic field increases, higher LBs become depopulated as they move through $E_F$ and their contributions to the Hall conductivity are removed. The Fermi energy will always intersect some finite DOS, resulting in a finite longitudinal resistivity. The flatness of a QQHE plateau is dependent on the specifics of the $k_z$ dispersion as well as disorder. We further note that phenomena such as (2D) edge modes, localization, QH scaling relations, and new correlated states may still be observable in 3D QQHE systems, offering a new route for investigating 3D QHE physics.

Semimetals are at the forefront of QQHE investigations as high mobilities imply $\rho_{xy}\approx \sigma_{xy}^{-1}$ and well defined LBs while low carrier densities enable experimental access to the regime where only the lowest few LBs are occupied. We demonstrate this, for the first time in \cdas{}, with a theoretical calculation of QL magnetotransport in Figs. \ref{fig:fig1s1}e,f: with $E_F\sim 20$ meV, clear quasi-quantized plateaus emerge in \rxy{} with accompanying features in \rxx{} (Methods and Supplementary Note 1). To our knowledge, these are the first calculations in \cdas{} In this work we wish to highlight an additional experimental reality in the pursuit of QQHE in low-density semimetals: while some amount of Coulomb disorder is necessary for observation of QQHE, too much disorder works to obscure the QQHE. As magnetotransport is one of the primary characterizations of QQHE systems, it is necessary to understand the ways in which charged disorder can affect these signatures. \cdas{} bulk-like thin films are a compelling platform for exploring the intersection of 3D QQHE and the magnetotransport effects of disorder because of the tunability of the carrier density. In \cdas{}, Coulomb disorder originates from native charged point defects like Cd vacancies and interstitials \cite{Brooks2023,Nelson2023}. These charged defects become screened, which results in a disorder potential $eV$ throughout the sample (Fig. 2a, Supplementary Note 5). The average size of this disorder potential, $eV_0$, in DSMs depends upon the Fermi level: As $E_F$ is reduced $eV_0$ grows because of reduced screening \cite{Skinner2014}. The size of this disorder potential relative to the Fermi level has been predicted to have dramatic influences on magnetotransport, summarized in Fig. 2b in \cdas{} \cite{Song2015,Rodionov:2023}. In the weak Coulomb disorder limit ($eV_0\ll E_F$), scattering from this potential generates the ubiquitous non-saturating linear MR in many TSMs \cite{Song2015,Leahy2018,Nelson2023}. In the limit of strong Coulomb disorder ($E_F\lesssim eV_0$), it has been predicted to result in quadratic MR \cite{Rodionov:2023}. Coulomb disorder will contribute to the non-QQHE magnetotransport, reduce mobilities, and broaden LBs. This effects of this disorder are represented as blue traces in Figs. \ref{fig:fig1s1}e,f. In the lowest density TSMs, where QQHE might be most experimentally accessible, Coulomb disorder may completely obscure QQHE magnetotransport and may be a reason QQHE has not been observed more often.

\section{Results}

\begin{figure*}[ht!]
    \centering
    \includegraphics[trim={.00cm 0cm 0cm 0cm},width = .9\linewidth,clip]{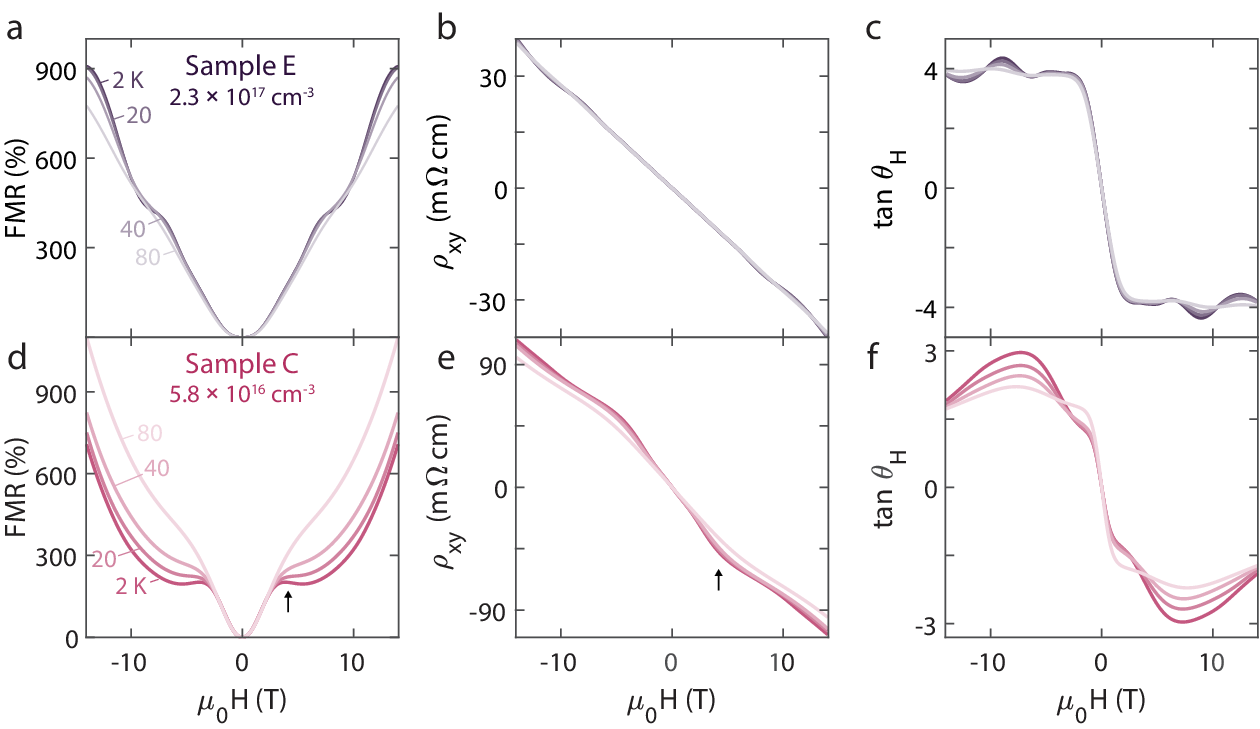}
    \caption{{\bf Carrier density tunes magnetotransport in \cdas.} a,d) Fractional magnetoresistance, b,e) Hall resistivity, and c,f) tangent of the Hall angle $\tan\theta_H\equiv\rho_{xy}/\rho_{xx}$ at several temperatures for a \cdas{} films with low temperature electron carrier densities of (a-c) Sample E: $2.3\times10^{17}$ \cm{} and (d-f) Sample C: $5.8\times10^{16}$ \cm{}. Sample E shows linear MR and Hall effect with superimposed quantum oscillations. Meanwhile, $\tan\theta_H$ saturates to a constant value. On the other hand, Sample C shows a low field quadratic and high field super-quadratic MR and a clear kink in \rxy{} (coincident features highlighted with arrows).   }
    \label{fig:Fig2}
\end{figure*}

We begin by exploring the 2 K magnetotransport in two \cdas{} films with differing carrier densities (mobilities and densities in Fig. \ref{fig:fig1s2}c). Fig. \ref{fig:Fig2}a-c are obtained on Sample E, a high carrier density film with $n_{2\textrm{K}} = 2.3\times 10^{17}$ \cm{}. The fractional magnetoresistance (FMR$=[\rho_{xx}(H)-\rho_0]/\rho_0)\times100\%$, where $\rho_0\equiv\rho_{xx}(0\textrm{T})$) scales linearly with increasing field with superimposed quantum oscillations (Fig. \ref{fig:Fig2}a). The Hall resistivity, \rxy{}, is linear with visible quantum oscillations, consistent with a single, electron-like carrier (Fig. \ref{fig:Fig2}b). Lastly, the tangent of the Hall angle, \tth{} $\equiv \rho_{xy}/\rho_{xx}$, rapidly increases at low fields and saturates to a constant, field-independent value, less any quantum oscillations (Fig. \ref{fig:Fig2}c). This behavior is standard and has been observed in many TSMs \cite{Nelson2023,Leahy2018}. 

We now turn to Sample C, shown in Fig. 2d-f, having $n_{2\textrm{K}}=5.8\times 10^{16}$\cm{}. The 2 K FMR (Fig. \ref{fig:Fig2}d) shows quadratic growth near zero field, followed by a shoulder above which the FMR continues to increase  rapidly and non-linearly. Remnants of this structure are visible in data up to 80 K. Correspondingly, \rxy{} exhibits a peculiar and distinct change in slope at a similar field which remains visible to higher temperatures (Fig. \ref{fig:Fig2}e).  Finally, \tth{} shares the usual low-field increase, but does not saturate as in the high-density sample, exhibiting a large lump near 5 T reflecting the structure of the underlying resistivities. 

The change in slope of the Hall resistivity and shouldered \rxx{} is striking. Such magnetotransport could originate from: i) multiband transport, ii) Berry curvature induced effects, or iii) quantum oscillations and QQHE. A two-band model can generate non-linear Hall effect and quadratic magnetoresistance, but multiple (or anisotropic) bands are necessary to generate a magnetoresistance with a rapidly changing magnetic field dependence. Nonmonotonicity is typically not observed in multiband model magnetoresistance, and nonmonotonicity in the Hall effect is typically smooth as a function of field, contrasting with the distinct slope change observed here. With a single band crossing the Fermi level, a multiband picture is difficult to justify. In the study of many Weyl semimetals, sharp slope changes in \rxy{} have been correlated with anomalous-like Hall effects originating from non-trivial, divergenceful Berry curvature originating from k-space-separated Weyl points. Applying a field to \cdas{}, the Dirac point splits into two sets of Weyl points separated in k-space along the applied field direction, resulting in nontrivial Berry curvature \cite{Ness:2025}. This splitting is very small at  our experimental fields and the effects on transport are likely minimal. 

The temperature and carrier density dependence of this behavior hints at its origins (Supplementary Note 4, Fig. S4). With decreasing carrier density, the features become more distinct and their position moves to lower fields. Increasing temperatures cause the \rxx{} and \rxy{} features to fade. As we show in the Supplement, these features evolve from oscillations, but the low carrier densities and near quantum limit fields generate the distinct dependence.

Recently, similar magnetotransport to that of Sample C has been observed in the zirconium and hafnium pentatellurides \cite{Tang2019,Galeski2020,Galeski2021,Piva2024}. Investigations in \zrte{} culminated in the realization of the 3D QQHE. Much like the 2D QHE, the QQHE exhibits peaks in \rxx{} paired with plateau-like features in \rxy{}. Correspondingly, \tth{} will approach some large value with periodic decreases between Hall plateaus. These QQHE dependencies transform from standard SdH oscillations when the lowest few LBs are occupied and sufficiently separated (Supplementary Note 4). Details of the 3D dispersion and other material parameters determine the nature of \rxx{} and \rxy{} plateau quasi-quantization.  The experimental observation of 'clean' \rxy{} plateaus and \rxx{} bumps in 3D metallic or semimetallic systems depends upon both low carrier concentrations (small Fermi surfaces) and sizeable mobilities. In \zrte{} and \hfte{}, low carrier densities ($0.9$-$2\times10^{17}$ \cm{}) paired with high mobilities are naturally accompanied by a $\tan\theta_H\sim10$ in the QL. Low carrier concentrations ensure that QL-sized fields can be attained with experimental ease (1-2 T) and high mobilities signal less LB broadening, yielding clearer QQHE features and ensuring $\rho_{xy} \sim \sigma_{xy}^{-1}$ \cite{Galeski2021,Tang2019}. As in the 2D QHE, the longitudinal and Hall resistivities have been observed to be related such that $\rho_{xx}=-\alpha H \partial_H\rho_{xy}$ \cite{Halperin1982,Halperin1987,Galeski2021,Gooth2023}. Such resistivity scaling relations originate from disorder and are not universal features of QHE yet are almost exclusively observed in 2D QHE and recently 3D QQHE systems \cite{Simon1994,Wawrzynczak2022,Galeski2021}. A 3D Fermi surface is necessary to distinguish QQHE from a quasi-2D QHE. In contrast to the case of the pentatellurides, we observe far fewer quantum oscillations in our \cdas{} films. The low carrier density and small Fermi surface size imply a low oscillation frequency. The field range where quantum oscillations are visible is further limited by the comparatively low mobilities of our films (0.67-2.6 $\times 10^4$ cm${}^2$/Vs compared to 3-50 $\times 10^4$ cm${}^2$/Vs in \zrte{}) \cite{Tang2019,Galeski2021}. Spin splitting, LB broadening, and the shape of the Fermi surface ellipsoid in \cdas{} further obscure oscillations. Despite these complications, we are able to verify a 3D Fermi surface in films of a slightly higher carrier density (Supplementary Note 6). We have shown by calculation that QQHE is apparent in the weak-disorder limit of \cdas{} (Fig. \ref{fig:fig1s1}e,f). Given these predictions, the fact that only the lowest few LBs are occupied in our \cdas{} films, and the qualitative similarity of the band structure and transport to the pentatellurides, it is likely that the distinct features observed in Sample C originate from the QQHE, though disorder serves to obscure the effects.

Disorder in TSMs contributes to the longitudinal magnetoresistivity. In \cdas{}, Coulomb disorder in the form of charged Cadmium vacancies and interstitials is screened, resulting in a disorder potential landscape of average magnitude $eV_0$ (as in Fig. \ref{fig:fig1s2}a) \cite{Skinner2014}. $eV_0$ depends sensitively on the carrier density: smaller carrier densities result in less effective charged defect screening and larger $eV_0$ (theoretical calculation shown in Fig. \ref{fig:fig1s2}b, Supplementary Note 5). It has been theoretically predicted and experimentally verified that scattering from this disorder potential in the limit of weak disorder ($eV_0\ll E_F$) is responsible for the nonsaturating linear FMR and field-saturated \tth{} observed in several semimetals \cite{Song2015,Leahy2018,Nelson2023}. In the limit of strong Coulomb disorder ($eV_0\gtrsim E_F$), the magnetic field dependence of the magnetoresistivity has been predicted to range from $\propto H^2$ to $\propto H$ contingent on the relative magnitudes of the applied field and $eV_0$ \cite{Rodionov:2023}. By tuning the carrier density in our \cdas{} films, we anticipate the ability to tune between the weak-and-strong Coulomb disorder regimes, which should be partly reflected in $\rho_{xx}(H)$. Reduced carrier density yields larger $eV_0$ which  results in more LB broadening obscuring oscillation phenomena. Overall, we hypothesize that the magnetoresistivity in \cdas{} is a superposition of QQHE and Coulomb disorder scattering effects. Our theoretical calculations when including moderate disorder reflect this (Fig. \ref{fig:fig1s1}f).

\begin{figure*}[ht!]
    \centering
    \includegraphics[trim={.00cm 0cm 0cm 0cm},width =1\linewidth,clip]{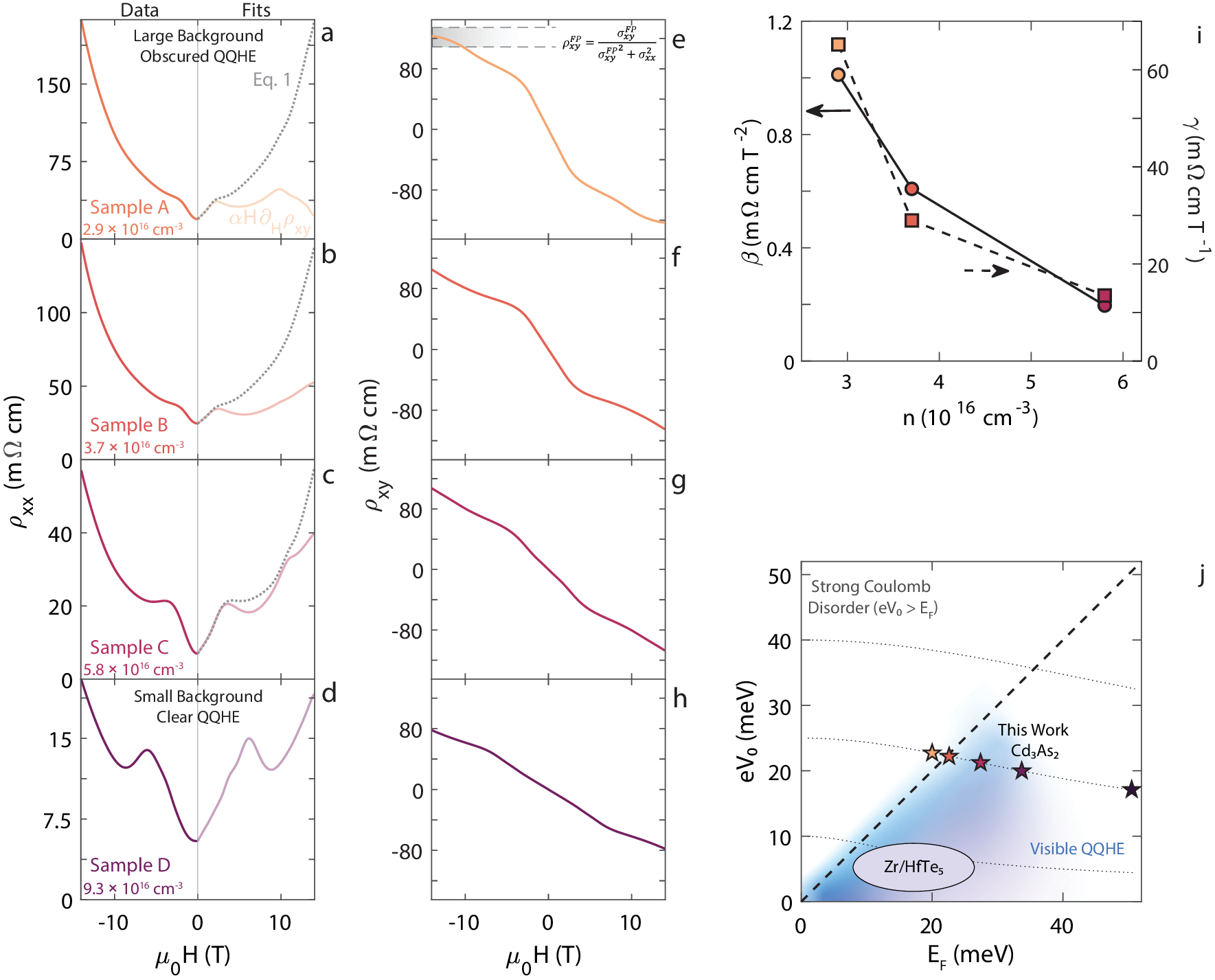}
    \caption{{\bf QQHE and Coulomb scattering in \cdas{}}. a-d) Longitudinal  and e-h) Hall resistivity for samples of increasing carrier density at 5 K. In panels a-d, the left half of the axes correspond to raw data while the right half show fits of the data to only the QQHE (light color, solid) and to the full Eq. 1 (gray, dotted). Eq. 1 was not fit to Sample D as it lies in the weak Coulomb disorder limit. In panel e, the quasi-quantized value of the Hall resistivity in the quantum limit is displayed, showing good agreement with experiment (calculation in Supplementary Note 7).  i) Carrier density dependence of $\beta$ and $\gamma$ fit parameters from Eq. 1. j) A plot of the disorder potential strength calculation vs. the Fermi level for \cdas{}. Significant Coulomb disorder in \cdas{} combats a clean observation of the QQHE. Meanwhile, for \zrte{} and \hfte{}, stronger screening of disorder gives weaker contributions from Coulomb disorder scattering. Estimates for the pentatelluride systems from \cite{Tang2019,Galeski2021,WangEvidence2018,GourgoutFO2022,Shahi2018,Galeski2020,Piva2024}. Dotted lines represent calculated contours of how $eV_0$ changes with Fermi level from Ref. \cite{Skinner2014} (Supplementary Note 5).  }
    \label{fig:Fig3}
\end{figure*}

We turn our attention to the magnetotransport for additional samples with varying carrier densities. Decreasing the carrier density simultaneously reduces the field scales of the observed QQHE and increases the effective Coulomb disorder strength. Figure \ref{fig:Fig3} shows \rxx{} (a-d) as dark colored lines for $\mu_0H<0$ T and \rxy{} (e-h). Sample carrier densities increase from top to bottom (panel a to panel d). As expected with decreasing carrier density, the slope of the Hall effect and the longitudinal resistivity increase in magnitude. While the step-like slope change in \rxy{} is visible for each sample, the character of \rxx{} changes significantly as a function of $n$. A high field \rxy{} plateau is observed in Sample A originating from the QQHE in the quantum limit. The plateau value is in agreement with the expected value based on theoretical calculations (Supplementary Note 7). In Sample D (Fig. \ref{fig:Fig3}d), the clear lump in \rxx{} at $\sim 5.7$ T coincides with the reduced slope in \rxy{}. To demonstrate the relative contributions of the QQHE and the background magnetoresistance, we plot an estimated QQHE contribution, $\alpha H\partial_H\rho_{xy}$, as light colored traces for $\mu_0H>0$ T in Fig. \ref{fig:Fig3}a-d. The overall proportionality constant, $\alpha$, is chosen to match the size of the lump-like feature in \rxx{}. Clearly, the size of the QQHE signal in Fig. \ref{fig:Fig3}d is large compared to the non-QQHE background as $\rho_{xx}$ is nearly identical to $\alpha H\partial_H\rho_{xy}$. For Sample D, this simple scaling works remarkably well. With decreasing $n$, this lone scaling progressively worsens: while the correlations of the features in \rxx{} and \rxy{} originating from the QQHE hold, an additional background component grows serving to obfuscate the \rxx{} QQHE features. Simultaneously, LBs are increasingly broadened as the carrier density is reduced and the Coulomb disorder strength grows (Supplemental Note 4). 

We hypothesize that the non-QQHE \rxx{} background and its apparent dependence on the carrier density originate from Coulomb disorder.  At high carrier densities, \cdas{} is in the weak Coulomb disorder limit, $eV_0<E_F$, accompanied by linear MR (Sample E). As the carrier density (and requisitely $E_F$) shrink, the magnitude of the average disorder potential increases because of reduced screening. Shrinking $E_F$ and increasing $eV_0$ move the system from weak to strong Coulomb disorder, with $eV_0\gtrsim E_F$ (Supplementary Note 5). After Ref. \cite{Rodionov:2023}, the strong Coulomb disorder magnetoresistance is expected to behave as $\sim H^2$ in the low field limit. With increasing field, the dependence crosses over to $\sim H\log\delta\sqrt{H}$. We construct a simple empirical model to describe our longitudinal resistivity that combines the QQHE resistivity scaling with the expected magnetoresistance in the limit of strong Coulomb disorder:
\begin{multline}
    \rho_{xx}-\rho_{0} = \alpha H\partial_H\rho_{xy} + f(H,H_c,w)\beta H^2 + \\ f(H,H_c,-w)\gamma H \log\delta\sqrt{H},
    \label{eq:Eq1}
\end{multline}
where $\rho_0$ is the zero field resistivity, $\alpha$ scales the QQHE, and $\beta$ and $\gamma$ are fitting parameters that scale the quadratic and sub-quadratic strong Coulomb disorder magnetoresistivities respectively, and $\delta$ is a fitting parameter related to the disorder strength and the Fermi velocity. $f(H,H_c,w)=(\exp{((H-H_c)/w)} + 1)^{-1}$ is a simple Fermi function that serves to smoothly crossover between different regimes of the strong Coulomb disorder magnetoresistivity. We fix $H_c = 9.6$ T and $w=1.5$ T, serving as global values for all samples (Supplementary Note 3). Fitting to equation \ref{eq:Eq1} is performed only for Samples A-C, as the strong Coulomb disorder magnetoresistance breaks down with increasing carrier density. In Supplementary Note 2, we discuss the regimes of validity of this empirical expression. 

Fits of Equation \ref{eq:Eq1} to \rxx{} are shown as gray dashed lines in Fig. \ref{fig:Fig3}a-c, demonstrating good agreement for all samples. Fig. \ref{fig:Fig3}i shows $\beta$ and $\gamma$ as a function of carrier density (a table of all fitting parameters can be found in Supplementary Note 3). We find similar values of $\delta\sim0.32$ T${}^{-1/2}$ for Samples A-C. The QQHE scaling factor lies between $\alpha\sim0.17-0.35$ for all samples. Consistent with our expectations from the raw $\rho_{xx}(H)$, $\beta$ and $\gamma$ decrease with increasing $n$ as the disorder effects become weaker with increased screening (additional details on the temperature dependencies of \rxx{} and \rxy{} can be found in Supplementary Note 4). At the same time, the QQHE component comprises a larger portion of the measured \rxx{}. This is consistent with a crossover from strong-to-weak Coulomb disorder with increasing carrier density. In the weak Coulomb disorder limit (Sample D), the magnetoresistance scales linearly with field and the 2D QHE scaling relation, $\alpha H\partial_H\rho_{xy}$, captures the QQHE and the effects of weak Coulomb disorder simultaneously. We highlight that while the 2D QHE resistivity scaling relation seems to hold in \cdas{}, it may differ in other QQHE systems.

\cdas{} occupies a unique space in the search for 3D QQHE materials. Systems with small carrier densities are the route to observe QQHE and MBE affords us this tunability in \cdas{}. At the same time, shrinking $E_F$ increases the effects of Coulomb disorder in \cdas{}: the average disorder potential increases, charge puddling smears the Fermi level throughout, Landau bands are broadened, and importantly, scattering from the screened Coulomb disorder changes character as $eV_0/E_F$ grows. Fig. \ref{fig:Fig3}j summarizes this effect, showing that \cdas{} crosses into a regime of strong Coulomb disorder with decreasing Fermi level. As we have hypothesized and shown, the contribution from the Coulomb disorder scattering grows relative to the QQHE contribution with decreasing carrier density. Cumulatively, the effects of larger $eV_0$ with decreasing $E_F$ conspire to partially obstruct clean QQHE in \cdas{}. A balance must be reached: while low $n$ enables the observation of QQHE in \cdas{} in the first place, it simultaneously reduces screening of charged defects, which in turn obscures the QQHE.

\section{Discussion} 

It is first insightful to distinguish the QQHE and magnetotransport observed in our films from other experimentally observed quantum Hall effects in \cdas{}. Clean 2D QHE has been observed in confined ultrathin films of \cdas{} grown by PLD and MBE \cite{Uchida2017,Schumann2018}. It has been proposed that this 2D QHE originates from a gapping of the bulk band structure with decreasing thickness, and softening of the quantization has been observed as a function of thickness, potentially consistent with a crossover to 3D QQHE behavior. All of our films are well within the bulk-like thickness regime ($> 100$ nm) and oscillation analyses where possible exhibit 3D Fermi surfaces, ruling out a 2D QHE (see Methods, Supplementary Note 6). Unique transport and surface quantum oscillations have been predicted in Dirac and Weyl semimetals originating from Weyl orbits \cite{Potter2014}. Surface quantum oscillations and 2D QH-like effects have been observed in \cdas{} material with thicknesses less than 100 nm, where bulk states may be gapped. These studies appear in conflict with earlier reports of 2D QHE in ultrathin films where confinement plays a role \cite{Moll2016,Uchida2017,Zhang2017,Zhang2019}. Regardless, the orientation and thickness of our films preclude the Weyl orbit picture.

Though some reports on thin film \cdas{} have observed similar magnetotransport to that shown here \cite{Nakazawa2019,Healhofer2020}, our results represent the first study of its kind to understand the experimental low carrier density magnetotransport as an amalgamation of a 3D QQHE and Coulomb disorder. In fact, it is Coulomb disorder that rules the roost of magnetotransport as the Fermi level is brought close to the Dirac point. Our study positions \cdas{} as a platform for exploring the 3D QQHE in a {\it gapless} Dirac system in the presence of tunable disorder. Our description intuitively extrapolates from the theoretically and experimentally motivated presence of significant Coulomb disorder in Dirac systems like \cdas{} and Na${}_3$Bi \cite{Skinner2014,Song2015,Edmonds2017,Leahy2018,Nelson2023}. The recent emergence of the 3D QQHE has brought an understanding that quasi-quantized \rxy{} and the corresponding \rxx{} lumps (an extension of quantum oscillations) are the rule -- not the exception -- in low carrier density systems \cite{Tang2019,Gooth2023}. We anticipate increasing the film mobility and minimizing the effects of Coulomb disorder are avenues for enhancing the QQHE in \cdas{}.

It is instructive to consider our framework in the context of the pentatelluride systems. Plateaus and peaks much sharper than those presented here have been observed in \zrte{} \cite{Tang2019,Galeski2021}. Where is the Coulomb disorder background in this case? We anticipate that charged defects in \zrte{} and \cdas{} are screened in the same manner \cite{Skinner2014}. A larger dielectric constant in \zrte{} \cite{GourgoutFO2022} helps to more effectively screen charged point defects and reduce $eV_0$. Additionally, different growth conditions in \zrte{} can alter the observed Coulomb disorder. A recent STM study on chemical vapor transport (CVT) and flux-grown crystals revealed significant charge puddling only in CVT-grown samples, which tend to have more Te vacancies resulting in more charged defects \cite{Salzmann2020}. Most literature reports on QQHE or 3D QHE in \zrte{} are obtained on flux-grown crystals. The signatures of QQHE in CVT grown crystals are slightly obscured as compared to their flux grown counterparts, likely indicative of higher charged defect concentrations (and corequisitely larger carrier densities) \cite{WangEvidence2018,GourgoutFO2022,Shahi2018}. Our study implies that this difference in part stems from differences in Coulomb disorder in these crystals. As a result of lower charged point defect concentrations and greater screening, we estimate a disorder potential of a few meV in the pentatelluride systems as shown in Fig. \ref{fig:Fig3}j. Overall, this reduced disorder potential helps place the pentatellurides in a regime where QQHE is more observable than in \cdas{} because of less LB broadening.

Broadly speaking, efforts to study 3D QQHE in TSMs and low-density metals will confront the realities of Coulomb disorder. We anticipate the QQHE to be more easily observable in systems with less prevalent charged disorder. We highlight that strategies that only reduce the Fermi level in TSMs will be met with the knock-on effects of reduced screening of charged defects: stronger Coulomb disorder scattering and broadening of LBs. Disorder must be simultaneously and independently addressed. Unlike bulk, thin film semimetals like those presented here offer tunability to realistically control and manipulate charge disorder. Passivation of charged point defects may be a promising route towards enhancing 3D QQHE. In the case of \cdas{}, gating experiments in bulk-like films will allow direct tuning of the Coulomb disorder strength in single samples, helping to understand these mechanisms. Additionally, further optimization of growth or post-growth annealing may be another route to alter charged disorder. The exploration of 3D QQHE and QHE phases awaits an understanding of how to manipulate and tune this disorder. 

\section{Summary}
In summary, we have observed evidence of the 3D QQHE in low carrier density, (001)-oriented \cdas{} thin films.  Our picture identifies Coulomb disorder as the common thread linking  magnetotransport behavior across the spectrum of varying carrier density and $E_F$ in \cdas{}, helping to unite a broad swath of observed magnetotransport reported in the literature. By reducing the carrier density in our films, we have probed the effects of the transition from weak-to-strong Coulomb disorder. With decreasing carrier density, the strength of Coulomb disorder increases, leading to the observation of strong Coulomb disorder magnetoresistivity in concert with the 3D QQHE.  Coulomb disorder is centrally influential in TSMs and serves to conflate clear QL transport in \cdas{}. This work will be undoubtedly useful in the burgeoning studies exploring QQHE in low-density metals and semimetals as well as the ongoing efforts to understand the effects of disorder as topological semimetals transition from fundamental curiosities to application. 

\section{Funding Declaration}
This work was authored in part by the National Laboratory of the Rockies for U.S. Department of energy (DOE) under Contract No. DE-AC36-08GO28308. 

Funding was provided by the U.S. Department of Energy, Office of Science, Basic Energy Sciences, Division of Materials Sciences and Engineering, Physical Behavior of Materials Program under the Disorder in Topological Semimetals project. 

The work at Sandia is supported by the LDRD program. 

The National High Magnetic Field Laboratory is supported by the National Science Foundation through NSF/DMR-2128556 and the State of Florida. 

The views expressed in the article do not necessarily represent the views of the DOE or the U.S. Government. The U.S. Government retains and the publisher, by accepting the article for publication, acknowledges that the U.S. Government retains a nonexclusive, paid-up, irrevocable, worldwide license to publish or reproduce the published form of this work, or allow others to do so, for U.S. Government purposes.

\section{Acknowledgements}

We would like to thank Igor \u Zuti\'c, Konstantin Denisov, and Rebecca Smaha for useful discussions. 

\section{Declaration of Interests}
The authors declare no competing interests.

\section{Methods}
{\bf Film Growth}
Films were grown as described elsewhere on GaAs(001) substrates \cite{Rice2022}. Cd$_3$As$_2$ layers were grown from separate Cd and As effusion sources on lattice matched Zn$_x$Cd${}_{1-x}$Te buffers using As rich conditions at 115 C. All samples have bulk-like thicknesses, ranging from 200-560 nm. 

{\bf Electrical Transport}
\cdas{} Hall bars with 6 electroplated Au contacts were fabricated using standard wet photolithography. Electrical transport measurements were performed by the current-reversal technique using a Keithley 2182A Nanovoltmeter with a 6221A AC and DC current source in a Quantum Design Physical Property Measurement System and by the low frequency AC technique using SRS lockin SRS860 with a Keithley 6221 AC and DC current source in a 30 T resistive magnet.  Carrier densities are extracted from the low field ($\mu_0H < 1$ T slopes of $\rho_{xy}$).

{\bf Theoretical Model}
We model the electronic structure of bulk Cd$_3$As$_2$ by using a $k\cdot p$ approach based on 4 orbitals ($s,p_{x,y,z}$) and two spins \cite{Wang:2013}. The corresponding Hamiltonian also contains a spin-orbit coupling term \cite{Wang:2013} as well a Zeeman term. The Hamiltonian (of size $8\times 8$) is a function of three parameters, i.e. the 3 k-space vectors $k_x,k_y,k_z$. Such a $k \cdot p$ model faithfully reproduces the ab-initio electronic structure of \cdas{}. Details of the model and first-principle calculation can be found in \cite{Ness2025}.

In order to take into account the LBs created by the application of an external magnetic field (along the z-direction to mimic the experimental conditions of our (001)-oriented films), we apply the so-called Peierls transformation and quantize the parameters $k_x,k_y$ in the plane perpendicular to the field into non-commuting operators \cite{Luttinger:1955,Harper:1991}. The latter are themselves expressed in term of creation and annihilation operators ($a^\dag$ and $a$) of the harmonic-oscillator-like states associated with the Landau bands (labeled by the integer number $n_\text{LB}$) \cite{Jeon:2014,Miao:2024,Smith:2024}. Upon the transformation $H(k_x,k_y,k_z)\rightarrow H(a,a^\dag;k_z)$, the new Hamiltonian is parameterized by only one wave-vector, i.e. $k_z$ parallel to the direction of the applied field. The new Hamiltonian has now a size of $(8\times (n_\text{LB}^\text{max} +1))^2$ where $n_\text{LB}^\text{max}$ corresponds to the maximum number of LBs used in the calculations.

The Hamiltonian is diagonalized for each value of the parameter $k_z$ to give the eigenvalues $E_{m,n_\text{LB}}(k_z)$ and eigenstates $\vert\Psi_{m,n_\text{LB}}(k_z)\rangle$, where $m$ labels the original $sp$ plus spin orbitals, $n_\text{LB}$ labels the different LB and $k_z$ is the only left continuous parameter. From these eigenstates and from the velocity matrices $\nabla_{k_\alpha}H(k_x,k_y,k_z)$ (transformed into the LB basis after $\nabla_k$ derivatives are taken), we can calculate the conductivity $\sigma_{\alpha\beta}$ ($\alpha,\beta=x,y,z$) within linear response theory \cite{Bastin:1971,Crepieux:2001,Morimoto:2012}:
\begin{align*}
\sigma_{\alpha\beta} &=
-i e^2 \hbar \sum_{\bar n,\bar m} f_{E_{\bar n}}\cdot
\\ &\left[
\frac{\langle \bar n\vert j_\alpha\vert \bar m \rangle \langle \bar m\vert j_\beta\vert \bar n\rangle}{(E_{\bar n} - E_{\bar m} -i\eta)^2} 
- 
\frac{\langle \bar n\vert j_\beta\vert \bar m \rangle \langle \bar m\vert j_\alpha\vert \bar n\rangle}{(E_{\bar n} - E_{\bar m} +i\eta)^2}
\right]
\label{eq:sigmaDCab}
\end{align*}

where $f_{E}$ is the Fermi function, and $\bar{m}$ and $\bar{n}$ are condensed labels which correspond to $m$, $n_\textrm{LB}$, and $k_z$ indices of the eigenstates. The Hall conductivity $\sigma_{xy}$ is obtained from the Berry curvature of the states as described in \cite{Gradhand:2012}:
\begin{equation*}
    \sigma_{\alpha\beta}=-\frac{e^2}{\hbar}\sum_{\bar{n}}f_{E_{\bar{n}}}\Omega_{\bar{n}},
\end{equation*}
where $\Omega_{\bar{n}}$ is the Berry curvature defined as:
\begin{equation*}
\begin{split}
\Omega_{\bar n}
& = i\hbar^2 \sum_{\bar m \ne \bar n} 
\frac{\langle \bar n\vert j_\alpha\vert \bar m \rangle \langle \bar m\vert j_\beta\vert \bar n\rangle 
- \langle \bar n\vert j_\beta\vert \bar m \rangle \langle \bar m\vert j_\alpha\vert \bar n\rangle}{(E_{\bar n} - E_{\bar m})^2} 
\end{split}
\label{eq:Bcurv_n}
\end{equation*}
To get a final conductivity in the above equations, an integration over $k_z$ is performed. For all our calculations of the electronic structure and the conductivities, we have carefully checked the convergence of the result versus the maximum number of LB $n_\textrm{LB}^\text{max}$ and the $k_z$ mesh used. For the longitudinal conductivity $\sigma_{\alpha\alpha}$, we keep the tiny imaginary part $i\eta$ in the energy denominators as it is numerically more convenient. Having this imaginary part is also a good starting point for the inclusion of weak disorder effects, as explained in Supplementary Note 1. Additionally, Supplementary Note 1 details more clearly the approximated manner in which we deal with moderate disorder effects in the calculations. 

Calculations of the conductivity have been performed for the cases of fixed chemical potential and fixed charge density. The longitudinal resistivity is calculated from the standard expression $\rho_{xx} = \sigma_{xx}/(\sigma_{xx}^2 + \sigma_{xy}^2)$. Note that from the symmetry of the Hamiltonian and a magnetic field applied in the $z$-direction, one has $\sigma_{xx}=\sigma_{yy}$ and $\sigma_{xy}=-\sigma_{yx}$. In the weak-disorder case, we always find $\sigma_{xy} \gg \sigma_{xx}$, hence $\rho_{xx} \sim \sigma_{xx}/\sigma_{xy}^2$
and $\rho_{xy} = \sigma_{xy}/(\sigma_{xx}^2 + \sigma_{xy}^2) \sim 1/\sigma_{xy}$.

Weak disorder calculations at fixed chemical potential produce, as expected, the "rounded" plateaux in the Hall conductivity versus applied field which is characteristic of the 3D QQHE, accompanied with peak-like  structure in the longitudinal conductivity located at the onset of the plateaux in the Hall conductivity (black traces shown in Figs. 1e,f). All these features correspond to magnetic field values for which a LB crosses the Fermi level of the Dirac/Weyl semimetal. For fixed charge density calculations, the 3D QQHE plateaux disappear and one recovers the (quasi)classical (i.e. not quantized) dependence of the Hall conductivity on the applied field, as shown in Supplementary Note 1.

\section{Author Contributions}
I. L., A. R., J. N., H. N., M. v. S., D. G., A. S., W. P., and K. A. contributed to the preparation of the manuscript. K. A., I. L., J. N., and A. R. conceived and designed experiments. I. L., J. N., A. R., D. G., and A. S. performed the experiments. I. L. and H. N. analyzed the data. I. L., J. N., H. N., A. R., and M. v. S. contributed materials or analysis tools. H. N. and M. v. S. performed the conductivity calculations. 

\section{Supplemental Information} 
The supplement gives more detailed information about 1) the theoretical $k\cdot p$ model and the inclusion of disorder in theoretical calculations of the magnetotransport, 2) the strong Coulomb disorder model, 3) the strong Coulomb disorder fitting parameters, 4) additional temperature dependent magnetotransport, 5) the calculation of the Coulomb disorder strength in \cdas{}, 6) analysis of quantum oscillations, and 7) the calculation of the Hall resistivity plateau value. 

\section{Data Availability}
The data that support the findings of this study are available from the corresponding author upon reasonable request.

\bibliography{Cd3As2_001.bib}

@article{Galeski2021,
	title = {Origin of the quasi-quantized {Hall} effect in {ZrTe5}},
	volume = {12},
	issn = {2041-1723},
	url = {https://www.nature.com/articles/s41467-021-23435-y},
	doi = {10.1038/s41467-021-23435-y},
	number = {1},
	urldate = {2024-10-09},
	journal = {Nature Communications},
	author = {Galeski, S. and Ehmcke, T. and Wawrzyńczak, R. and Lozano, P. M. and Cho, K. and Sharma, A. and Das, S. and Küster, F. and Sessi, P. and Brando, M. and Küchler, R. and Markou, A. and König, M. and Swekis, P. and Felser, C. and Sassa, Y. and Li, Q. and Gu, G. and Zimmermann, M. V. and Ivashko, O. and Gorbunov, D. I. and Zherlitsyn, S. and Förster, T. and Parkin, S. S. P. and Wosnitza, J. and Meng, T. and Gooth, J.},
	month = may,
	year = {2021},
	pages = {3197},
}

@article{Tang2019,
	title = {Three-dimensional quantum {Hall} effect and metal–insulator transition in {ZrTe5}},
	volume = {569},
	issn = {0028-0836, 1476-4687},
	url = {https://www.nature.com/articles/s41586-019-1180-9},
	doi = {10.1038/s41586-019-1180-9},
	number = {7757},
	urldate = {2024-10-09},
	journal = {Nature},
	author = {Tang, Fangdong and Ren, Yafei and Wang, Peipei and Zhong, Ruidan and Schneeloch, John and Yang, Shengyuan A. and Yang, Kun and Lee, Patrick A. and Gu, Genda and Qiao, Zhenhua and Zhang, Liyuan},
	month = may,
	year = {2019},
	pages = {537--541},
}

@article{Wang:2013,
  title = {Three-dimensional Dirac semimetal and quantum transport in Cd${}_{3}$As${}_{2}$},
  author = {Wang, Zhijun and Weng, Hongming and Wu, Quansheng and Dai, Xi and Fang, Zhong},
  journal = {Phys. Rev. B},
  volume = {88},
  issue = {12},
  pages = {125427},
  year = {2013},
  doi = {10.1103/PhysRevB.88.125427},
  url = {https://link.aps.org/doi/10.1103/PhysRevB.88.125427},
}

@article{Luttinger:1955,
  title = {Motion of Electrons and Holes in Perturbed Periodic Fields},
  author = {Luttinger, J. M. and Kohn, W.},
  journal = {Phys. Rev.},
  volume = {97},
  issue = {4},
  pages = {869--883},
  numpages = {0},
  year = {1955},
  month = {Feb},
  publisher = {American Physical Society},
  doi = {10.1103/PhysRev.97.869},
  url = {https://link.aps.org/doi/10.1103/PhysRev.97.869},
}

@article{Harper:1991,
doi = {10.1088/0953-8984/3/18/001},
url = {https://dx.doi.org/10.1088/0953-8984/3/18/001},
year = {1991},
volume = {3},
number = {18},
pages = {3047},
author = {P G Harper},
title = {Magnetic gauge transformations in solid-state problems},
journal = {Journal of Physics: Condensed Matter},
}

@Article{Jeon:2014,
author={Jeon, Sangjun
and Zhou, Brian B.
and Gyenis, Andras
and Feldman, Benjamin E.
and Kimchi, Itamar
and Potter, Andrew C.
and Gibson, Quinn D.
and Cava, Robert J.
and Vishwanath, Ashvin
and Yazdani, Ali},
title={Landau quantization and quasiparticle interference in the three-dimensional Dirac semimetal Cd3As2},
journal={Nature Materials},
year={2014},
volume={13},
number={9},
pages={851-856},
doi={10.1038/nmat4023},
url={https://doi.org/10.1038/nmat4023}
}

@article{Smith:2024,
  title = {Theory for ${\mathrm{Cd}}_{3}{\mathrm{As}}_{2}$ thin films in the presence of magnetic fields},
  author = {Smith, M. and Quito, Victor L. and Burkov, A. A. and Orth, P. P. and Martin, I.},
  journal = {Phys. Rev. B},
  volume = {109},
  issue = {15},
  pages = {155136},
  numpages = {16},
  year = {2024},
  month = {Apr},
  publisher = {American Physical Society},
  doi = {10.1103/PhysRevB.109.155136},
  url = {https://link.aps.org/doi/10.1103/PhysRevB.109.155136}
}

@article{Gradhand:2012,
doi = {10.1088/0953-8984/24/21/213202},
url = {https://dx.doi.org/10.1088/0953-8984/24/21/213202},
year = {2012},
volume = {24},
number = {21},
pages = {213202},
author = {M Gradhand and D V Fedorov and F Pientka and P Zahn and I Mertig and B L Györffy},
title = {First-principle calculations of the Berry curvature of Bloch states for charge and spin transport of electrons},
journal = {Journal of Physics: Condensed Matter},
}

@article{Miao:2024,
  title = {Engineering the in-plane anomalous Hall effect in ${\mathrm{Cd}}_{3}{\mathrm{As}}_{2}$ thin films},
  author = {Miao, Wangqian and Guo, Binghao and Stemmer, Susanne and Dai, Xi},
  year = {2024},
  journal = {Phys. Rev. B},
  volume = {109},
  issue = {15},
  pages = {155408},
  numpages = {8},
  doi = {10.1103/PhysRevB.109.155408},
  url = {https://link.aps.org/doi/10.1103/PhysRevB.109.155408}
}

@article{Bastin:1971,
author = {A. Bastin and C. Lewiner and O. Betbeder-Matibet and P. Nozieres},
title = {Quantum Oscillations of the {H}all effect of a {F}ermion gaz with random impurity scattering},
journal = {J. Phys. Chem. Solids},
volume = {32},
pages = {1811-1824},
year = {1971},
}

@article{Crepieux:2001,
  title = {Theory of the anomalous Hall effect from the Kubo formula and the Dirac equation},
  author = {Cr\'epieux, A. and Bruno, P.},
  journal = {Phys. Rev. B},
  volume = {64},
  pages = {014416},
  numpages = {16},
  year = {2001},
  doi = {10.1103/PhysRevB.64.014416},
  url = {https://link.aps.org/doi/10.1103/PhysRevB.64.014416}
}

@article{Morimoto:2012,
doi = {10.1088/1742-6596/400/4/042047},
url = {https://dx.doi.org/10.1088/1742-6596/400/4/042047},
year = {2012},
volume = {400},
number = {4},
pages = {042047},
author = {Takahiro Morimoto and Hideo Aoki},
title = {Flow diagram of the longitudinal and Hall conductivities in ac regime in the disordered graphene quantum Hall system},
journal = {Journal of Physics: Conference Series},
}

@article{Rodionov:2023,
  title = {Quantum magnetoresistance of Weyl semimetals with strong Coulomb disorder},
  author = {Rodionov, Ya. I. and Kugel, K. I. and Aronzon, B. A.},
  journal = {Phys. Rev. B},
  volume = {107},
  issue = {15},
  pages = {155120},
  year = {2023},
  doi = {10.1103/PhysRevB.107.155120},
  url = {https://link.aps.org/doi/10.1103/PhysRevB.107.155120}
}

@article{Skinner2014,
	title = {Coulomb disorder in three-dimensional {Dirac} systems},
	volume = {90},
	copyright = {http://link.aps.org/licenses/aps-default-license},
	issn = {1098-0121, 1550-235X},
	url = {https://link.aps.org/doi/10.1103/PhysRevB.90.060202},
	doi = {10.1103/PhysRevB.90.060202},
	number = {6},
	urldate = {2024-10-09},
	journal = {Physical Review B},
	author = {Skinner, Brian},
	month = aug,
	year = {2014},
	pages = {060202},
}

@article{Nelson2023,
	title = {Direct link between disorder and magnetoresistance in topological semimetals},
	volume = {107},
	issn = {2469-9950, 2469-9969},
	url = {https://link.aps.org/doi/10.1103/PhysRevB.107.L220206},
	doi = {10.1103/PhysRevB.107.L220206},
	number = {22},
	urldate = {2024-10-09},
	journal = {Physical Review B},
	author = {Nelson, Jocienne N. and Leahy, Ian A. and Rice, Anthony D. and Brooks, Chase and Teeter, Glenn and Van Schilfgaarde, Mark and Lany, Stephan and Fluegel, Brian and Lee, Minhyea and Alberi, Kirstin},
	month = jun,
	year = {2023},
	pages = {L220206},
}

@article{Brooks2023,
	title = {Band energy dependence of defect formation in the topological semimetal {Cd} 3 {As} 2},
	volume = {107},
	issn = {2469-9950, 2469-9969},
	url = {https://link.aps.org/doi/10.1103/PhysRevB.107.224110},
	doi = {10.1103/PhysRevB.107.224110},
	number = {22},
	urldate = {2024-10-09},
	journal = {Physical Review B},
	author = {Brooks, Chase and Van Schilfgaarde, Mark and Pashov, Dimitar and Nelson, Jocienne N. and Alberi, Kirstin and Dessau, Daniel S. and Lany, Stephan},
	month = jun,
	year = {2023},
	pages = {224110},
}

@article{Song2015,
	title = {Linear magnetoresistance in metals: {Guiding} center diffusion in a smooth random potential},
	volume = {92},
	copyright = {http://link.aps.org/licenses/aps-default-license},
	issn = {1098-0121, 1550-235X},
	shorttitle = {Linear magnetoresistance in metals},
	url = {https://link.aps.org/doi/10.1103/PhysRevB.92.180204},
	doi = {10.1103/PhysRevB.92.180204},
	number = {18},
	urldate = {2024-10-09},
	journal = {Physical Review B},
	author = {Song, Justin C. W. and Refael, Gil and Lee, Patrick A.},
	month = nov,
	year = {2015},
	pages = {180204},
}

@article{Edmonds2017,
	title = {Spatial charge inhomogeneity and defect states in topological {Dirac} semimetal thin films of {Na} $_{\textrm{3}}$ {Bi}},
	volume = {3},
	issn = {2375-2548},
	url = {https://www.science.org/doi/10.1126/sciadv.aao6661},
	doi = {10.1126/sciadv.aao6661},
	number = {12},
	urldate = {2024-10-09},
	journal = {Science Advances},
	author = {Edmonds, Mark T. and Collins, James L. and Hellerstedt, Jack and Yudhistira, Indra and Gomes, Lídia C. and Rodrigues, João N. B. and Adam, Shaffique and Fuhrer, Michael S.},
	month = dec,
	year = {2017},
	pages = {eaao6661},
}

@article{Gooth2023,
	title = {Quantum-{Hall} physics and three dimensions},
	volume = {86},
	issn = {0034-4885, 1361-6633},
	url = {https://iopscience.iop.org/article/10.1088/1361-6633/acb8c9},
	doi = {10.1088/1361-6633/acb8c9},
	number = {4},
	urldate = {2024-10-09},
	journal = {Reports on Progress in Physics},
	author = {Gooth, Johannes and Galeski, Stanislaw and Meng, Tobias},
	month = apr,
	year = {2023},
	pages = {044501},
}

@article{Klitzing1980,
	title = {New {Method} for {High}-{Accuracy} {Determination} of the {Fine}-{Structure} {Constant} {Based} on {Quantized} {Hall} {Resistance}},
	volume = {45},
	copyright = {http://link.aps.org/licenses/aps-default-license},
	issn = {0031-9007},
	url = {https://link.aps.org/doi/10.1103/PhysRevLett.45.494},
	doi = {10.1103/PhysRevLett.45.494},
	number = {6},
	urldate = {2024-10-10},
	journal = {Physical Review Letters},
	author = {Klitzing, K. V. and Dorda, G. and Pepper, M.},
	month = aug,
	year = {1980},
	pages = {494--497},
}

@article{Klitzing1986,
	title = {The quantized {Hall} effect},
	volume = {58},
	copyright = {http://link.aps.org/licenses/aps-default-license},
	issn = {0034-6861},
	url = {https://link.aps.org/doi/10.1103/RevModPhys.58.519},
	doi = {10.1103/RevModPhys.58.519},
	number = {3},
	urldate = {2024-10-10},
	journal = {Reviews of Modern Physics},
	author = {Von Klitzing, Klaus},
	month = jul,
	year = {1986},
	pages = {519--531},
}

@article{Wawrzynczak2022,
	title = {Quasi-quantized {Hall} response in bulk {InAs}},
	volume = {12},
	issn = {2045-2322},
	url = {https://www.nature.com/articles/s41598-022-05916-2},
	doi = {10.1038/s41598-022-05916-2},
	number = {1},
	urldate = {2024-10-10},
	journal = {Scientific Reports},
	author = {Wawrzyńczak, R. and Galeski, S. and Noky, J. and Sun, Y. and Felser, C. and Gooth, J.},
	month = feb,
	year = {2022},
	pages = {2153},
}

@article{Galeski2020,
	title = {Unconventional {Hall} response in the quantum limit of {HfTe5}},
	volume = {11},
	issn = {2041-1723},
	url = {https://www.nature.com/articles/s41467-020-19773-y},
	doi = {10.1038/s41467-020-19773-y},
	number = {1},
	urldate = {2024-10-10},
	journal = {Nature Communications},
	author = {Galeski, S. and Zhao, X. and Wawrzyńczak, R. and Meng, T. and Förster, T. and Lozano, P. M. and Honnali, S. and Lamba, N. and Ehmcke, T. and Markou, A. and Li., Q. and Gu, G. and Zhu, W. and Wosnitza, J. and Felser, C. and Chen, G. F. and Gooth, J.},
	month = nov,
	year = {2020},
	pages = {5926},
}

@article{Piva2024,
	title = {Importance of the semimetallic state for the quantum {Hall} effect in {HfTe} 5},
	volume = {8},
	issn = {2475-9953},
	url = {https://link.aps.org/doi/10.1103/PhysRevMaterials.8.L041202},
	doi = {10.1103/PhysRevMaterials.8.L041202},
	number = {4},
	urldate = {2024-10-10},
	journal = {Physical Review Materials},
	author = {Piva, M. M. and Wawrzyńczak, R. and Kumar, Nitesh and Kutelak, L. O. and Lombardi, G. A. and Dos Reis, R. D. and Felser, C. and Nicklas, M.},
	month = apr,
	year = {2024},
	pages = {L041202},
}

@article{Manna2022,
	title = {Three-dimensional quasiquantized {Hall} insulator phase in {Sr} {Si} 2},
	volume = {106},
	issn = {2469-9950, 2469-9969},
	url = {https://link.aps.org/doi/10.1103/PhysRevB.106.L041113},
	doi = {10.1103/PhysRevB.106.L041113},
	number = {4},
	urldate = {2024-10-10},
	journal = {Physical Review B},
	author = {Manna, K. and Kumar, N. and Chattopadhyay, S. and Noky, J. and Yao, M. and Park, J. and Förster, T. and Uhlarz, M. and Chakraborty, T. and Schwarze, B. V. and Hornung, J. and Strocov, V. N. and Borrmann, H. and Shekhar, C. and Sun, Y. and Wosnitza, J. and Felser, C. and Gooth, J.},
	month = jul,
	year = {2022},
	pages = {L041113},
}

@article{Zhang2017,
	title = {Evolution of {Weyl} orbit and quantum {Hall} effect in {Dirac} semimetal {Cd3As2}},
	volume = {8},
	issn = {2041-1723},
	url = {https://www.nature.com/articles/s41467-017-01438-y},
	doi = {10.1038/s41467-017-01438-y},
	number = {1},
	urldate = {2024-10-10},
	journal = {Nature Communications},
	author = {Zhang, Cheng and Narayan, Awadhesh and Lu, Shiheng and Zhang, Jinglei and Zhang, Huiqin and Ni, Zhuoliang and Yuan, Xiang and Liu, Yanwen and Park, Ju-Hyun and Zhang, Enze and Wang, Weiyi and Liu, Shanshan and Cheng, Long and Pi, Li and Sheng, Zhigao and Sanvito, Stefano and Xiu, Faxian},
	month = nov,
	year = {2017},
	pages = {1272},
}

@article{Nakazawa2019,
	title = {Molecular beam epitaxy of three-dimensionally thick {Dirac} semimetal {Cd3As2} films},
	volume = {7},
	issn = {2166-532X},
	url = {https://pubs.aip.org/apm/article/7/7/071109/122510/Molecular-beam-epitaxy-of-three-dimensionally},
	doi = {10.1063/1.5098529},
	number = {7},
	urldate = {2024-10-10},
	journal = {APL Materials},
	author = {Nakazawa, Y. and Uchida, M. and Nishihaya, S. and Sato, S. and Nakao, A. and Matsuno, J. and Kawasaki, M.},
	month = jul,
	year = {2019},
	pages = {071109},
}

@article{Zhang2019,
	title = {Quantum {Hall} effect based on {Weyl} orbits in {Cd3As2}},
	volume = {565},
	issn = {0028-0836, 1476-4687},
	url = {https://www.nature.com/articles/s41586-018-0798-3},
	doi = {10.1038/s41586-018-0798-3},
	number = {7739},
	urldate = {2024-10-10},
	journal = {Nature},
	author = {Zhang, Cheng and Zhang, Yi and Yuan, Xiang and Lu, Shiheng and Zhang, Jinglei and Narayan, Awadhesh and Liu, Yanwen and Zhang, Huiqin and Ni, Zhuoliang and Liu, Ran and Choi, Eun Sang and Suslov, Alexey and Sanvito, Stefano and Pi, Li and Lu, Hai-Zhou and Potter, Andrew C. and Xiu, Faxian},
	month = jan,
	year = {2019},
	pages = {331--336},
}

@article{Leahy2018,
	title = {Nonsaturating large magnetoresistance in semimetals},
	volume = {115},
	issn = {0027-8424, 1091-6490},
	url = {https://pnas.org/doi/full/10.1073/pnas.1808747115},
	doi = {10.1073/pnas.1808747115},
	number = {42},
	urldate = {2024-10-10},
	journal = {Proceedings of the National Academy of Sciences},
	author = {Leahy, Ian A. and Lin, Yu-Ping and Siegfried, Peter E. and Treglia, Andrew C. and Song, Justin C. W. and Nandkishore, Rahul M. and Lee, Minhyea},
	month = oct,
	year = {2018},
	pages = {10570--10575},
}

@article{Salzmann2020,
	title = {Nature of native atomic defects in {ZrTe} 5 and their impact on the low-energy electronic structure},
	volume = {4},
	issn = {2475-9953},
	url = {https://link.aps.org/doi/10.1103/PhysRevMaterials.4.114201},
	doi = {10.1103/PhysRevMaterials.4.114201},
	number = {11},
	urldate = {2024-10-10},
	journal = {Physical Review Materials},
	author = {Salzmann, B. and Pulkkinen, A. and Hildebrand, B. and Jaouen, T. and Zhang, S. N. and Martino, E. and Li, Q. and Gu, G. and Berger, H. and Yazyev, O. V. and Akrap, A. and Monney, C.},
	month = nov,
	year = {2020},
	pages = {114201},
}

@article{WangEvidence2018,
	title = {Evidence for {Layered} {Quantized} {Transport} in {Dirac} {Semimetal} {ZrTe5}},
	volume = {8},
	issn = {2045-2322},
	url = {https://www.nature.com/articles/s41598-018-23011-3},
	doi = {10.1038/s41598-018-23011-3},
	number = {1},
	urldate = {2024-10-10},
	journal = {Scientific Reports},
	author = {Wang, Wei and Zhang, Xiaoqian and Xu, Huanfeng and Zhao, Yafei and Zou, Wenqin and He, Liang and Xu, Yongbing},
	month = mar,
	year = {2018},
	pages = {5125},
}

@article{GourgoutFO2022,
	title = {Magnetic freeze-out and anomalous {Hall} effect in {ZrTe5}},
	volume = {7},
	issn = {2397-4648},
	url = {https://www.nature.com/articles/s41535-022-00478-y},
	doi = {10.1038/s41535-022-00478-y},
	number = {1},
	urldate = {2024-10-10},
	journal = {npj Quantum Materials},
	author = {Gourgout, Adrien and Leroux, Maxime and Smirr, Jean-Loup and Massoudzadegan, Maxime and Lobo, Ricardo P. S. M. and Vignolles, David and Proust, Cyril and Berger, Helmuth and Li, Qiang and Gu, Genda and Homes, Christopher C. and Akrap, Ana and Fauqué, Benoît},
	month = jul,
	year = {2022},
	pages = {71},
}

@article{Shahi2018,
	title = {Bipolar {Conduction} as the {Possible} {Origin} of the {Electronic} {Transition} in {Pentatellurides}: {Metallic} vs {Semiconducting} {Behavior}},
	volume = {8},
	issn = {2160-3308},
	shorttitle = {Bipolar {Conduction} as the {Possible} {Origin} of the {Electronic} {Transition} in {Pentatellurides}},
	url = {https://link.aps.org/doi/10.1103/PhysRevX.8.021055},
	doi = {10.1103/PhysRevX.8.021055},
	number = {2},
	urldate = {2024-10-10},
	journal = {Physical Review X},
	author = {Shahi, P. and Singh, D.J. and Sun, J.P. and Zhao, L.X. and Chen, G.F. and Lv, Y.Y. and Li, J. and Yan, J.-Q. and Mandrus, D.G. and Cheng, J.-G.},
	month = may,
	year = {2018},
	pages = {021055},
}

@article{Halperin1982,
	title = {Quantized {Hall} conductance, current-carrying edge states, and the existence of extended states in a two-dimensional disordered potential},
	volume = {25},
	copyright = {http://link.aps.org/licenses/aps-default-license},
	issn = {0163-1829},
	url = {https://link.aps.org/doi/10.1103/PhysRevB.25.2185},
	doi = {10.1103/PhysRevB.25.2185},
	number = {4},
	urldate = {2024-10-10},
	journal = {Physical Review B},
	author = {Halperin, B. I.},
	month = feb,
	year = {1982},
	pages = {2185--2190},
}

@article{Laughlin1981,
	title = {Quantized {Hall} conductivity in two dimensions},
	volume = {23},
	copyright = {http://link.aps.org/licenses/aps-default-license},
	issn = {0163-1829},
	url = {https://link.aps.org/doi/10.1103/PhysRevB.23.5632},
	doi = {10.1103/PhysRevB.23.5632},
	number = {10},
	urldate = {2024-10-10},
	journal = {Physical Review B},
	author = {Laughlin, R. B.},
	month = may,
	year = {1981},
	pages = {5632--5633},
}

@article{Cooper1993,
	title = {Coulomb interactions and the integer quantum {Hall} effect: {Screening} and transport},
	volume = {48},
	copyright = {http://link.aps.org/licenses/aps-default-license},
	issn = {0163-1829, 1095-3795},
	shorttitle = {Coulomb interactions and the integer quantum {Hall} effect},
	url = {https://link.aps.org/doi/10.1103/PhysRevB.48.4530},
	doi = {10.1103/PhysRevB.48.4530},
	number = {7},
	urldate = {2024-10-10},
	journal = {Physical Review B},
	author = {Cooper, N. R. and Chalker, J. T.},
	month = aug,
	year = {1993},
	pages = {4530--4544},
}

@article{Stormer1998,
	title = {Nobel {Lecture}: {The} fractional quantum {Hall} effect},
	volume = {71},
	copyright = {http://link.aps.org/licenses/aps-default-license},
	issn = {0034-6861, 1539-0756},
	shorttitle = {Nobel {Lecture}},
	url = {https://link.aps.org/doi/10.1103/RevModPhys.71.875},
	doi = {10.1103/RevModPhys.71.875},
	number = {4},
	urldate = {2024-10-10},
	journal = {Reviews of Modern Physics},
	author = {Stormer, Horst L.},
	month = jul,
	year = {1999},
	pages = {875--889},
}

@article{Yu2018,
	title = {Pi and 4*Pi Josephson Effects Mediated by a Dirac Semimetal},
	volume = {120},
	issn = {0031-9007, 1079-7114},
	url = {https://link.aps.org/doi/10.1103/PhysRevLett.120.177704},
	doi = {10.1103/PhysRevLett.120.177704},
	number = {17},
	urldate = {2024-11-24},
	journal = {Physical Review Letters},
	author = {Yu, W. and Pan, W. and Medlin, D.L. and Rodriguez, M.A. and Lee, S.R. and Bao, Zhi-qiang and Zhang, F.},
	month = apr,
	year = {2018},
	pages = {177704},
}

@article{LiSM2018,
	title = {Spin-momentum locking and spin-orbit torques in magnetic nano-heterojunctions composed of {Weyl} semimetal {WTe2}},
	volume = {9},
	issn = {2041-1723},
	url = {https://www.nature.com/articles/s41467-018-06518-1},
	doi = {10.1038/s41467-018-06518-1},
	number = {1},
	urldate = {2024-11-24},
	journal = {Nature Communications},
	author = {Li, Peng and Wu, Weikang and Wen, Yan and Zhang, Chenhui and Zhang, Junwei and Zhang, Senfu and Yu, Zhiming and Yang, Shengyuan A. and Manchon, A. and Zhang, Xi-xiang},
	month = sep,
	year = {2018},
	pages = {3990},
}

@article{Rice2022,
	title = {Epitaxial {Dirac} {Semimetal} {Vertical} {Heterostructures} for {Advanced} {Device} {Architectures}},
	volume = {32},
	issn = {1616-301X, 1616-3028},
	url = {https://onlinelibrary.wiley.com/doi/10.1002/adfm.202111470},
	doi = {10.1002/adfm.202111470},
	number = {21},
	urldate = {2024-11-25},
	journal = {Advanced Functional Materials},
	author = {Rice, Anthony D. and Lee, Choong Hee and Fluegel, Brian and Norman, Andrew G. and Nelson, Jocienne N. and Jiang, Chun Sheng and Steger, Mark and McGott, Deborah L. and Walker, Patrick and Alberi, Kirstin},
	month = may,
	year = {2022},
	pages = {2111470},
}

@article{Kohmoto1992,
	title = {Diophantine equation for the three-dimensional quantum {Hall} effect},
	volume = {45},
	copyright = {http://link.aps.org/licenses/aps-default-license},
	issn = {0163-1829, 1095-3795},
	url = {https://link.aps.org/doi/10.1103/PhysRevB.45.13488},
	doi = {10.1103/PhysRevB.45.13488},
	number = {23},
	urldate = {2025-06-02},
	journal = {Physical Review B},
	author = {Kohmoto, Mahito and Halperin, Bertrand I. and Wu, Yong-Shi},
	month = jun,
	year = {1992},
	pages = {13488--13493},
}

@article{Halperin1987,
	title = {Possible {States} for a {Three}-{Dimensional} {Electron} {Gas} in a {Strong} {Magnetic} {Field}},
	volume = {26},
	issn = {0021-4922, 1347-4065},
	url = {https://iopscience.iop.org/article/10.7567/JJAPS.26S3.1913},
	doi = {10.7567/JJAPS.26S3.1913},
	number = {S3-3},
	urldate = {2025-06-03},
	journal = {Japanese Journal of Applied Physics},
	author = {Halperin, Bertrand I.},
	month = jan,
	year = {1987},
	pages = {1913},
}

@article{Healhofer2020,
	title = {Topological {Insulator} {State} and {Collapse} of the {Quantum} {Hall} {Effect} in a {Three}-{Dimensional} {Dirac} {Semimetal} {Heterojunction}},
	volume = {10},
	issn = {2160-3308},
	url = {https://link.aps.org/doi/10.1103/PhysRevX.10.011050},
	doi = {10.1103/PhysRevX.10.011050},
	number = {1},
	urldate = {2025-06-06},
	journal = {Physical Review X},
	author = {Kealhofer, David A. and Galletti, Luca and Schumann, Timo and Suslov, Alexey and Stemmer, Susanne},
	month = feb,
	year = {2020},
	pages = {011050},
}

@article{Potter2014,
	title = {Quantum oscillations from surface {Fermi} arcs in {Weyl} and {Dirac} semimetals},
	volume = {5},
	issn = {2041-1723},
	url = {https://www.nature.com/articles/ncomms6161},
	doi = {10.1038/ncomms6161},
	number = {1},
	urldate = {2025-06-06},
	journal = {Nature Communications},
	author = {Potter, Andrew C. and Kimchi, Itamar and Vishwanath, Ashvin},
	month = oct,
	year = {2014},
	pages = {5161},
}

@article{Moll2016,
	title = {Transport evidence for {Fermi}-arc-mediated chirality transfer in the {Dirac} semimetal {Cd3As2}},
	volume = {535},
	issn = {0028-0836, 1476-4687},
	url = {https://www.nature.com/articles/nature18276},
	doi = {10.1038/nature18276},
	number = {7611},
	urldate = {2025-06-10},
	journal = {Nature},
	author = {Moll, Philip J. W. and Nair, Nityan L. and Helm, Toni and Potter, Andrew C. and Kimchi, Itamar and Vishwanath, Ashvin and Analytis, James G.},
	month = jul,
	year = {2016},
	pages = {266--270},
}

@article{Uchida2017,
	title = {Quantum {Hall} states observed in thin films of {Dirac} semimetal {Cd3As2}},
	volume = {8},
	issn = {2041-1723},
	url = {https://www.nature.com/articles/s41467-017-02423-1},
	doi = {10.1038/s41467-017-02423-1},
	number = {1},
	urldate = {2025-06-10},
	journal = {Nature Communications},
	author = {Uchida, Masaki and Nakazawa, Yusuke and Nishihaya, Shinichi and Akiba, Kazuto and Kriener, Markus and Kozuka, Yusuke and Miyake, Atsushi and Taguchi, Yasujiro and Tokunaga, Masashi and Nagaosa, Naoto and Tokura, Yoshinori and Kawasaki, Masashi},
	month = dec,
	year = {2017},
	pages = {2274},
}

@article{Simon1994,
	title = {Explanation for the {Resistivity} {Law} in {Quantum} {Hall} {Systems}},
	volume = {73},
	copyright = {http://link.aps.org/licenses/aps-default-license},
	issn = {0031-9007},
	url = {https://link.aps.org/doi/10.1103/PhysRevLett.73.3278},
	doi = {10.1103/PhysRevLett.73.3278},
	number = {24},
	urldate = {2025-06-17},
	journal = {Physical Review Letters},
	author = {Simon, Steven H. and Halperin, Bertrand I.},
	month = dec,
	year = {1994},
	pages = {3278--3281},
}

@misc{Ness:2025,
      author={H. Ness and I. Leahy and A. Rice and D. Pashov and K. Alberi and M. van Schilfgaarde},
      title={Doping {T}opological {D}irac {S}emimetal with magnetic impurities: electronic structure of {M}n-doped {C}d$_3${A}s$_2$}, 
      year={2025},
      eprint={2505.06662},
      archivePrefix={arXiv},
      url={https://arxiv.org/abs/2505.06662}, 
}

@article{Koshino2003,
	title = {Integer quantum {Hall} effect in isotropic three-dimensional crystals},
	volume = {67},
	copyright = {http://link.aps.org/licenses/aps-default-license},
	issn = {0163-1829, 1095-3795},
	url = {https://link.aps.org/doi/10.1103/PhysRevB.67.195336},
	doi = {10.1103/physrevb.67.195336},
	number = {19},
	urldate = {2025-07-10},
	journal = {Physical Review B},
	author = {Koshino, M. and Aoki, H.},
	year = {2003},
	note = {Publisher: American Physical Society (APS)},
}

@article{Koshino2001,
	title = {Hofstadter {Butterfly} and {Integer} {Quantum} {Hall} {Effect} in {Three} {Dimensions}},
	volume = {86},
	copyright = {http://link.aps.org/licenses/aps-default-license},
	issn = {0031-9007, 1079-7114},
	url = {https://link.aps.org/doi/10.1103/PhysRevLett.86.1062},
	doi = {10.1103/physrevlett.86.1062},
	number = {6},
	urldate = {2025-07-10},
	journal = {Physical Review Letters},
	author = {Koshino, M. and Aoki, H. and Kuroki, K. and Kagoshima, S. and Osada, T.},
	year = {2001},
	note = {Publisher: American Physical Society (APS)},
	pages = {1062--1065},
}

@article{Cooper1989,
	title = {Quantized {Hall} effect and a new field-induced phase transition in the organic superconductor ({TMTSF}){2PF6}},
	volume = {63},
	copyright = {http://link.aps.org/licenses/aps-default-license},
	issn = {0031-9007},
	url = {https://link.aps.org/doi/10.1103/PhysRevLett.63.1984},
	doi = {10.1103/physrevlett.63.1984},
	number = {18},
	urldate = {2025-07-10},
	journal = {Physical Review Letters},
	author = {Cooper, J. R. and Kang, W. and Auban, P. and Montambaux, G. and Jérome, D. and Bechgaard, K.},
	year = {1989},
	note = {Publisher: American Physical Society (APS)},
	pages = {1984--1987},
}

@article{Schumann2018,
	title = {Observation of the {Quantum} {Hall} {Effect} in {Confined} {Films} of the {Three}-{Dimensional} {Dirac} {Semimetal} {Cd3As2}},
	volume = {120},
	copyright = {https://link.aps.org/licenses/aps-default-license},
	issn = {0031-9007, 1079-7114},
	url = {https://link.aps.org/doi/10.1103/PhysRevLett.120.016801},
	doi = {10.1103/physrevlett.120.016801},
	number = {1},
	urldate = {2025-07-10},
	journal = {Physical Review Letters},
	author = {Schumann, Timo and Galletti, Luca and Kealhofer, David A. and Kim, Honggyu and Goyal, Manik and Stemmer, Susanne},
	year = {2018},
	note = {Publisher: American Physical Society (APS)},
}

@article{Ness2025, 
title={Doping topological Dirac semimetal with magnetic impurities: Electronic structure of Mn-doped Cd${}_3$As${}_2$}, 
volume={7}, 
ISSN={2643-1564}, 
DOI={10.1103/ww7d-qsg9},
number={3}, 
journal={Physical Review Research}, 
author={Ness, H. and Leahy, I. A and Rice, A. D. and Pashov, D. and Alberi, K. and Van Schilfgaarde, M.}, year={2025}, pages={033261} }

\end{document}